# Hydrogenation of accreting C-atoms and CO molecules - simulating ketene and acetaldehyde formation under dark and translucent cloud conditions –

Gleb Fedoseev[1,2], Danna Qasim[1,*], Ko-Ju Chuang[1], Sergio Ioppolo[3], Thanja Lamberts[4,1], Ewine F. van Dishoeck[5], and Harold Linnartz[1]

**ABSTRACT**

Simple and complex organic molecules (COMs) are observed along different phases of star and planet formation and have been successfully identified in prestellar environments such as dark and translucent clouds. Yet the picture of organic molecule formation at those earliest stages of star formation is not complete and an important reason is the lack of specific laboratory experiments that simulate carbon atom addition reactions on icy surfaces of interstellar grains. Here we present experiments in which CO molecules as well as C- and H-atoms are co-deposited with $H_2O$ molecules on a 10 K surface mimicking the ongoing formation of an "$H_2O$-rich" ice mantle. To simulate the effect of impacting C-atoms and resulting surface reactions with ice components, a specialized C-atom beam source is used, implemented on SURFRESIDE$^3$, an UHV cryogenic setup. Formation of ketene ($CH_2CO$) in the solid state is observed "in situ" by means of reflection absorption IR spectroscopy. $C^{18}O$ and D isotope labelled experiments are performed to further validate the formation of ketene. Data analysis supports that $CH_2CO$ is formed through C-atom addition to a CO-molecule, followed by successive hydrogenation transferring the formed :CCO into ketene. Efficient formation of ketene is in line with the absence of an activation barrier in C+CO reaction reported in the literature. We also discuss and provide experimental evidence for the formation of acetaldehyde ($CH_3CHO$) and possible formation of ethanol ($CH_3CH_2OH$), two COM derivatives of $CH_2CO$ hydrogenation. The underlying reaction network is presented and the astrochemical implications of the derived pathways are discussed.

[1] Laboratory for Astrophysics, Leiden Observatory, Leiden University, PO Box 9513, 2300 RA Leiden, the Netherlands
[2] Research Laboratory for Astrochemistry, Ural Federal University, Kuibysheva St. 48, 620026 Ekaterinburg, Russia
[3] School of Electronic Engineering and Computer Science, Queen Mary University of London, Mile End Road, London E1 4NS, UK
[4] Leiden Institute of Chemistry, Leiden University, Einsteinweg 55, 2333 CC, Leiden, the Netherlands
[5] Leiden Observatory, Leiden University, P.O. Box 9513, NL 2300 RA Leiden, the Netherlands
[*] Current address: Astrochemistry Laboratory, NASA Goddard Space Flight Center, Greenbelt, MD 20771, USA

# 1. INTRODUCTION

Astronomical gas-phase observations show that a variety of simple and complex organic molecules (COMs) exist in translucent and dense clouds as well as dark cores (Matthews et al. 1985; Turner et al. 1999; Bacmann et al. 2012; Soma et al. 2018). Astrochemical models show that gas-phase reactions generally lack the efficiency to explain the observed COM abundances (Millar et al. 1991; Charnley et al. 1992; Garrod et al. 2006). For this reason, it is assumed that many of the observed COMs are primarily formed on the icy surfaces of inter- (and circum)stellar dust grains, that act as cryogenic molecule reservoirs and catalytic surfaces.

The onset of solid-state COM formation is usually associated with the beginning of the CO freeze-out stage in dense clouds, where gas-phase atomic carbon is largely locked-up in CO (see van Dishoeck & Black 1988, Snow & McCall 2006). CO forms an ice coating layer on top of a previously formed $H_2O$-rich ice (Pontoppidan 2006, Boogert et al. 2015) and is hydrogenated by accreting H-atoms to form $H_2CO$, $CH_3OH$, and various ≥2 carbon bearing COMs, such as aldehydes, aldoses and (poly)ols. Indeed, laboratory experiments, as well as computational calculations, have demonstrated that the carbon backbone of COMs can be formed through $CH_nO$ radical recombinations on the surface and in the bulk of interstellar ices (Woods et al. 2012, Fedoseev et al. 2015, Butscher et al. 2015, 2017, Chuang et al. 2016, Alvarez-Barcia et al. 2018, Soma et al. 2018, Lamberts et al. 2019). These findings have been taken into account in astrochemical models to help explain the presence and abundances of various COMs in interstellar clouds (Woods et al. 2012, Vasyunin et al. 2017, Simons et al. 2020, Jin et al. 2020). The CO-ice coating is continuously exposed to various energetic particles, such as cosmic rays and cosmic ray induced UV photons, resulting in the formation of various excited species including ions and radicals that can react and further increase the chemical complexity in the ice (Öberg et al. 2009, Modica & Palumbo 2010, Maity et al. 2015, Paardekooper et al. 2016, Fedoseev et al. 2017, Chuang et al. 2017).

Following astronomical detections of COMs in dark and translucent clouds, recent laboratory experiments and computational efforts attempted to move the onset of efficient COM formation to an earlier stage, prior to or along with the formation of the $H_2O$-rich ice layer in the interstellar medium (Hudson & Loeffler 2013, Qasim et al. 2019; Chuang et al. 2020, 2021, Ioppolo et al. 2021). In these regions, atoms such as H, O, C, and N freeze-out onto bare dust grains (van de Hulst 1946; Boogert et al. 2015), where they can react to form COMs along with the formation of already identified $H_2O$, $CO_2$, $CH_4$, and $NH_3$.

Following this specific scenario, the growth of the carbon skeleton takes place through direct C-atom addition to or insertion in the various C-carbon bearing radicals produced along with simpler species (Tielens & Charnley 1997; Charnley 2001; Charnley & Rodgers 2005). The key reaction step responsible for the growth of the carbon skeleton and formation of 2-carbon bearing COMs suggested in these modeling works is the insertion of a C-atom into the HĊO radical, ultimately resulting in the formation of ketene ($CH_2CO$):

$$CO + H \rightarrow H\dot{C}O, \quad (1a)$$
$$H\dot{C}O + C \rightarrow H\dot{C}CO, \quad (1b)$$
$$H\dot{C}CO + H \rightarrow CH_2CO. \quad (1c)$$

Strictly speaking, ketene is not a COM because of the astronomical definition that COMs comprise of at least six atoms. However, ketene is an important oxygen-bearing organic compound that acts as a COM precursor due to the presence of unsaturated CC and CO bonds. The produced $CH_2CO$ can be further hydrogenated to form the astronomically defined COM derivatives, acetaldehyde ($CH_2CO + 2H \rightarrow CH_3CHO$) and ethanol ($CH_3CHO + 2H \rightarrow CH_3CH_2OH$), see Bisschop et al (2007).

An alternative route to form $CH_2CO$ was recently found in experiments condensing atomic carbon together with $H_2$, $H_2O$, and CO molecules inside a superfluid helium nanodroplet (Krasnokutski et al. 2017). The key reaction step responsible for the growth of the carbon skeleton in this scenario is the association of CO molecules with methylene:

$$H_2 + C \rightarrow H_2C:, \quad (2a)$$
$$H_2C: + CO \rightarrow CH_2CO. \quad (2b)$$

The second reaction step (2b) does, however, have a 1500 – 1700 K barrier that needs to be overcome (King et al. 1998; Ogihara et al. 2010). This may decrease the overall feasibility of this reaction route in view of the competing reactions with H atoms and molecular $H_2$.

Recent quantum chemical calculations offer a third reaction path that can lead to ketene formation in the ice following a barrierless association reaction between CO molecules and C atoms, see Papakondylis & Mavridis 2019:

$$CO + C \rightarrow :CCO, \quad (3a)$$
$$:CCO + H \rightarrow H\dot{C}CO, \quad (3b)$$
$$H\dot{C}CO + H \rightarrow CH_2CO \quad (3c)$$

where (3c) is identical to (1c).

To date, there are no reported experimental studies that have tested such carbon-atom induced reaction pathways on the surface of $H_2O$-rich interstellar ice analogues, i.e., using (more) representative astronomical conditions. The topic of the present study is to experimentally investigate the formation of ketene following the third suggested mechanism as well as the formation of two of its proposed derivatives, acetaldehyde and ethanol, under solid state conditions resembling the early phases of dark clouds or translucent clouds in which atom addition reactions play a key role. In the next sections, we discuss experiments in which C-atoms and H-atoms are co-deposited with CO molecules along with overabundance of $H_2O$ molecules. The latter simulates the growth of an interstellar $H_2O$-rich ice mantle *before* the CO accretion phase, i.e., the so-called CO freeze-out stage, when the solid-state chemistry is mainly driven by hydrogenation of CO molecules. The dominating reaction routes under these experimental conditions are proposed. In section 2, the experimental procedure is described. Section 3 presents the results. The paper is completed with a discussion on the astronomical implications, which incorporates the newly verified reaction pathways into a larger astrochemically relevant reaction network.

## 2. EXPERIMENTAL

The experiments are performed using SURFRESIDE[3], an ultra-high vacuum setup extended from a two-atom beam line system as described by Ioppolo et al. (2013) to a system with an additional C-atom beam source (Qasim et al. 2020b). In both papers details on the experimental procedures have been described in detail. Ices are grown on a gold-coated cupper substrate that is mounted on the cold finger of a closed-cycle He cryostat and placed in the centre of the main UHV chamber with a base pressure on the order of $10^{-10}$ mbar. Simultaneous co-deposition of $H_2O$ molecules with C atoms, CO molecules and H atoms is performed to grow a mixed ice analogue on the 10 K cold substrate. Isotope labelling using $C^{18}O$ isotope or D atoms is used for further validation.

A beam of H atoms mixed with undissociated $H_2$ molecules or D atoms mixed with undissociated $D_2$ molecules is obtained by partial dissociation of molecular $H_2$ or $D_2$ in a microwave discharge beam line (Oxford Scientific, Schmidt et al. 1996, Anton et al. 2000) mounted in a separate vacuum chamber with a base pressure of ~$1\times10^{-9}$ mbar. Any charged particles are deflected by means of an applied electric field which prohibits them from reaching the ice surface. Then, the electronically or vibrationally excited neutral species in the obtained beam are de-excited through collisions with the walls of an U-shaped quartz pipe placed along the path of the beam, and directed to the substrate plane. The used H-atom flux is derived to be $2\times10^{12}$ cm$^{-2}$ s$^{-1}$. This value is evaluated using a quadrupole mass spectrometer (QMS) with the ion head positioned on the place of the substrate along the path of the atom beam. See also Ioppolo et al.

(2013). First, the D-atom beam flux values are estimated for a range of relevant plasma parameters. Then the H-atom beam flux values are obtained by multiplication of the D-atom beam flux values by factor of 2 (rounded) accounting for the higher thermal velocity of the lighter H-atoms.

A C-atom beam with a $C_n/C$ (n>1) ratio of less than 0.01 is produced by a state-of-the-art customized C-atom source based on a commercial design (Dr. Eberl, MBE, Krasnokutski & Huisken 2014, Albar et al. 2017) located in another separate vacuum chamber with a base pressure of $(3-5) \times 10^{-9}$ mbar. A series of apertures are used to collimate the C-atom beam on the substrate avoiding deposition of carbon on the walls of the main UHV chamber. The C-atom flux used in this study is $\sim 5 \times 10^{11}$ cm$^{-2}$ s$^{-1}$, and only ground state C($^3$P) atoms are expected to be present in the beam. The C-atom flux is measured by performing a set of C+$^{18}$O$_2$ co-deposition experiments and quantifying the yield of this barrierless reaction, see Qasim et al. (2020b) for more details. The produced C-atom beam contains minor fractions of CO and CO$_2$ as main contaminants. A third separate vacuum chamber with a base pressure of $\sim 1 \times 10^{-9}$ mbar is used for H$_2$O molecular beam deposition with a rate equal to $8 \times 10^{12}$ cm$^{-2}$ s$^{-1}$. C$^{18}$O-gas is prepared in a regular all-metal dosing line prepumped to $<1 \times 10^{-4}$ mbar and guided towards the substrate through an all-metal leak valve and a capillary. The CO and C$^{18}$O deposition rates are determined to be $8 \times 10^{11}$ cm$^{-2}$ s$^{-1}$ and $1.4 \times 10^{12}$ cm$^{-2}$ s$^{-1}$ at the substrate surface, respectively.

The ice growth is monitored "*in situ*" by means of Reflection Absorption InfraRed Spectroscopy (RAIRS, Greenler 1966). In the present study, the band strength ratio between transmission and reflection modes is assumed to be constant for all studied species. The setup specific conversion factor between the band strength values of two modes equals 5.5 ± 0.8 and is obtained by comparing the values for the pure solid CO, H$_2$CO and CH$_3$OH reported for our setup geometry in Chuang et al. (2018) to the values reported for transmission in Bouilloud et al. (2015). Both measurements utilize a laser interference technique (at 632.8 nm laser wavelength) allowing to evaluate the absolute thickness of an ice with known refractive index by measuring the interference curve of the reflected light. Subsequently, the absolute thickness of the ice can be converted into an absorption band strength by using the ice density available from the literature.

Additional diagnostic information is extracted through temperature programmed desorption (TPD) and continuous monitoring of the gas-phase composition by means of a QMS. A 5 K/min heating rate with a data acquisition rate of one point per 0.5 K is used in the experiments. The combination of RAIRS and TPD-QMS, particularly when using isotopes, has been shown to provide an effective tool to realize unambiguous identifications of newly formed species and even to discriminate between structural isomers (Maity et al. 2015, Fedoseev et al. 2017, Chuang et al. 2020).

A significant overabundance of $H_2O$ molecules is used in the performed experiments. On one hand, this aims to represent the surface coverage during the 'H$_2$O-rich' ice formation stage of molecular clouds. On the other hand $H_2O$ also acts as a diluting buffer preventing direct reactions of deposited C-atoms or direct impact of depositing C-atoms on the various products formed in the reactions. The present work utilises about one-two orders of magnitude lower total H-atom fluences than those typically used in the previous CO+H experiments which aimed to address the events occurring later, during the CO freeze-out stage, see Fuchs et al. (2009), Fedoseev et al. (2015), Chuang et al. (2016).

## 3. RESULTS
### 3.1. IR spectroscopy

Ketene formation is observed in all experiments where CO molecules are co-deposited with C- and H- atoms and $H_2O$ molecules onto the 10 K substrate. The assignment of regular ketene ($^{12}CH_2^{12}C^{16}O$) through its $v_2$ band at 2133 cm$^{-1}$ (Arendale & Fletcher 1956, 1957) is not straightforward because of spectral overlap with the CO ice absorption feature (Hudson & Loeffler 2013). However, an unambiguous detection of newly formed $CH_2CO$ becomes possible by performing a temperature programmed desorption (TPD) experiment during RAIRS, as CO (~30 K) and $CH_2CO$ (~100 K) have very distinct desorption temperatures. The formation of ketene at a low temperature of 10 K, however, can only be confirmed through the use of isotopes such as D and, in particular, $^{18}O$ enriched $C^{18}O$. The use of a $C^{18}O$ isotope precursor allows to unambiguously discriminate among IR absorption features originating from $CH_2C^{18}O$ or $C^{18}O$ (Hudson & Loeffler 2013).

Figure 1 presents the RAIR spectra obtained for three selected experiments involving C-atom deposition with $H_2O$ molecules along with CO and H (spectrum a), CO and D (spectrum b) and $CO/C^{18}O$ and H (spectrum c). The experiments are performed for identical settings. The amount of deposited $H_2O$ is the same in all of the presented experiments, which is confirmed by comparing the absorbance of the broad water absorption band peaking around 1660 cm$^{-1}$ ($v_2$), (Gerakines et al. 1995). In the same way similar amounts of newly formed formaldehyde are found, based on the comparison of the absorbances of its characteristic features at 1718 cm$^{-1}$ ($v_2$), 1499 cm$^{-1}$ ($v_3$), and 1252 cm$^{-1}$ ($v_5$) (Hidaka et al. 2004, 2009). No efficient formation of $D_2CO$ or $H_2C^{18}O$ is observed in our experiments, as no clear absorption signals are found at 1676 cm$^{-1}$ for D$_2$CO ($v_2$) and/or 1694 cm$^{-1}$ ($v_2$) and 1488 cm$^{-1}$ ($v_3$) for H$_2$C$^{18}$O. This indicates that H$_2$CO in our experiments is primarily formed through reactions of C atoms with H$_2$O molecules, while the direct hydrogenation of CO molecules by H atoms is a minor channel under our selected experimental conditions. This is in line with the lower total H-atom fluence used in the present work in comparison with the values used in the previous CO+H experiments (Fuchs et al. 2009, Fedoseev et al. 2015, Chuang et al. 2016). Absorption features of CD$_4$ ($v_1$, 2251 cm$^{-1}$) and CH$_4$ ($v_3$, 1303 cm$^{-1}$) are, however, clearly observed in spectrum (b) and (c) of Fig. 1, respectively. This is consistent with the lack of an activation barrier for the hydrogenation of C atoms into methane as reported very recently (Qasim et al. 2020a), whereas CO hydrogenation involves an activation barrier (Watanabe & Kouchi 2002).

It should be noted, that the amount of H$_2$CO formed through interaction of C-atoms with H$_2$O molecules in our experiments scales directly with the C-atom flux and H$_2$O surface coverage, and is the same for the experiments performed at 10 and 25 K. The formation of H$_2$CO from C and H$_2$O is an important finding, and is subject of a follow-up study, currently pending for review. For the present work it is important to note that the fraction of C atoms that does not react with H$_2$O is assumed to be thermalized and available for the reaction with CO molecules and H atoms.

The multi-component absorption feature centred at 2137 cm$^{-1}$ with a high wavenumber component around 2150 cm$^{-1}$ in spectrum (a) of Fig. 1 is assigned to CO molecules with an unknown fraction of CH$_2$CO. The use of D-atoms, see spectrum (b) of Fig. 1, results in only a tentative detection of CD$_2$CO because of the weak intensity decrease of the peak centred at 2137 cm$^{-1}$ contributed by unlabelled CH$_2$CO and observation of the poorly resolved absorption features of CD$_2$CO at 2152 cm$^{-1}$ ($v_1$) and 2116 cm$^{-1}$ ($v_2$), see comparison between spectra (a) and (b) in the right panel of Fig. 1 (Arendale & Fletcher 1957, Kogure et al. 1992). Only in the experiment with C$^{18}$O addition, see spectrum (c) of Fig. 1, the CH$_2$C$^{18}$O absorption feature centred at 2107 cm$^{-1}$ ($v_2$, Hudson & Loeffler 2013) can be clearly resolved from the C$^{18}$O feature centred at 2087 cm$^{-1}$. The appearance of the CH$_2$C$^{18}$O absorption feature correlates with the visual decrease in the intensity of the CH$_4$ absorption feature centred at 1303 cm$^{-1}$ ($v_3$) in spectrum (c) of Fig. 1. This is in line with the appearance of a new C-atom consumption channel which competes with the production of CH$_4$ and yields CH$_2$C$^{18}$O in addition to the regular CH$_2$CO isotope.

To further validate this CH$_2$C$^{18}$O assignment, spectra are recorded after heating the ice to 55 K and 165 K. These spectra are shown in spectra (d) and

(e) of Fig. 1, respectively. At 55 K, i.e., well above the CO desorption temperature, only the small fraction of CO and C$^{18}$O entrapped in H$_2$O ice remains, making the distinct CH$_2$C$^{18}$O absorption feature even more evident. The spectrum at 165 K is obtained after desorption of H$_2$O ice together with all the entrapped highly volatile compounds. The characteristic bands around 2100 cm$^{-1}$ are all gone, confirming that the absorption at 2107 cm$^{-1}$ is not caused by unidentified low volatile species.

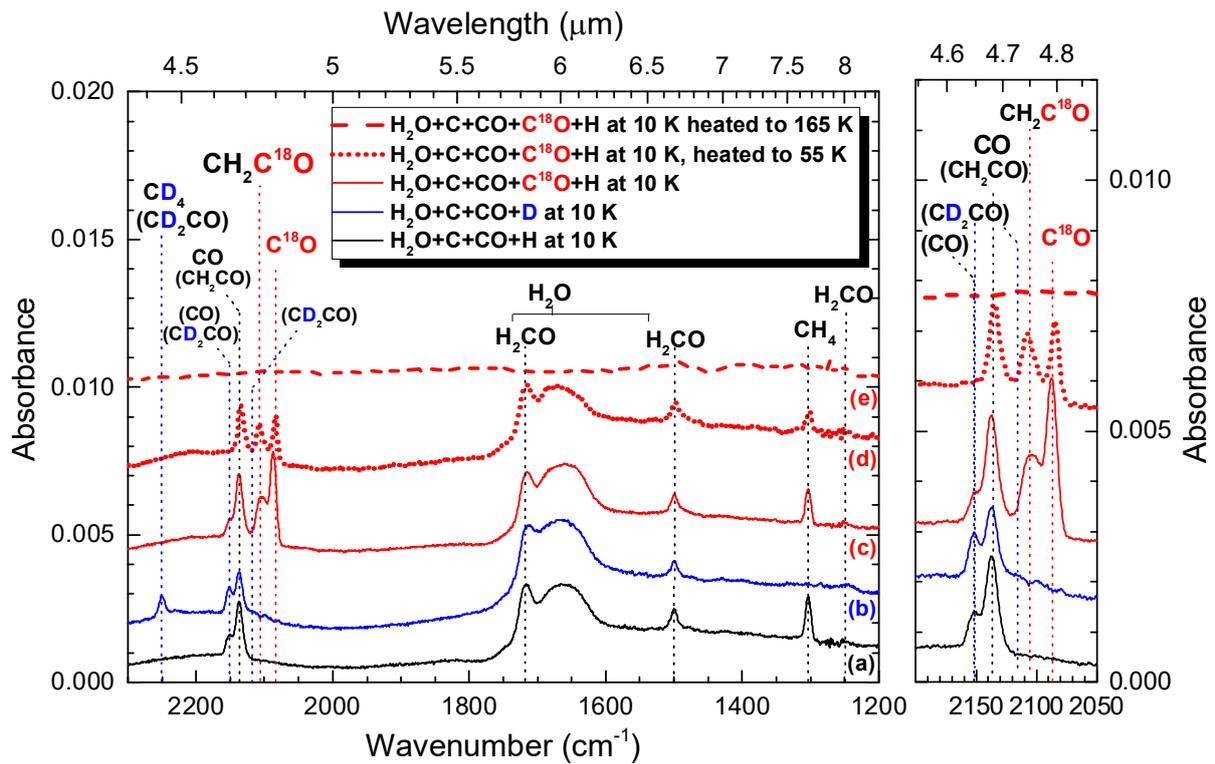

**Figure 1.** In the left panel the IR spectra are shown obtained after the simultaneous co-deposition of H$_2$O molecules with C-atoms and (a) CO molecules and H atoms; (b) CO molecules and D atoms; (c) a mixture of CO with C$^{18}$O molecules with H-atoms at a substrate temperature 10 K. All co-depositions are performed for a period of 1920 seconds. Spectra (d) and (e) are obtained after warm up of the ice presented in spectrum (c) to 55 K and 165 K, respectively. The H$_2$O:C:CO:(C$^{18}$O):H(D) ratio is estimated to be 16:1:1.5:(3):4(3) with the used C-atom flux equal to 5x10$^{11}$ cm$^{-2}$. The right panel shows a zoom-in of the 2200-2050 cm$^{-1}$ region where the main absorption features of regular and isotopic labelled CO and H$_2$CCO molecules are situated (note that the vertical axis is not identical to that used in the left panel). Spectra are offset for clarity. Blue and red

colours are used to indicate D and $^{18}$O labelled experiments and assignments, respectively. Tentative assignments are presented in brackets.

### 3.2. Mass spectrometry

In order to provide further evidence for the formation of ketene in the performed experiments, an example of a typical TPD QMS measurement is presented in Figure 2. This measurement is obtained for a co-deposition of CO molecules with C and H atoms and overabundance of $H_2O$ molecules. The four panels show the desorption curves for selected m/z values assigned to $H_2O$ (a), $CH_2CO$ (b), $CH_3CHO$ (c), and $CH_3CH_2OH$ (d). Ketene can be identified through its m/z signals at 14 ($CH_2^+$), 41 ($HCCO^+$), and 42 ($CH_2CO^+$) and reveals two desorption features (Fig. 2b). The weaker peak can be identified in the range from 88 to 107 K, and corresponds to the desorption temperature reported previously in the literature, see Maity et al. (2015), Chuang et al. (2020). The second feature in the range from 125 to 165 K corresponds to the fraction of entrapped ketene desorbing with $H_2O$ ice (see also Fig. 2a). Whereas in the IR spectra no resolved acetaldehyde and ethanol features could be unambiguously assigned, the TPD QMS shows that, along with the $CH_2CO$ itself, these two possible $CH_2CO$ hydrogenation products ($CH_3CHO$ and $CH_3CH_2OH$) can be identified as well. Fig. 2c shows $CH_3CHO$ signatures with characteristic m/z signals at 43 ($CH_2CHO^+$), 44 ($CH_3CHO^+$) and 15 ($CH_3^+$). Similarly to ketene, TPD data in Fig. 2c reveal two desorption features, i.e., the feature around a characteristic desorption temperature of $CH_3CHO$ at 120-140 K (Abplanalp et al. 2016) and a feature at 140-160 K coinciding with the desorption of the bulk of $H_2O$ ice. Fig. 2d shows possible features of $CH_3CH_2OH$ through (less prominent) mass peaks at m/z= 45 ($C_2H_5O^+$) and m/z=46 ($C_2H_5OH^+$) between 150 and 163 K. The thermal desorption of $CH_3CH_2OH$ has been reported to appear at temperatures several K higher than that of water ice, see Bergantini et al. (2017), Chuang et al. (2020). This finding is confirmed in Fig. 2.

No other COMs such as glycolaldehyde ($HOCH_2CHO$) or ethylene glycol ($HOCH_2CH_2OH$) are observed in the present work within the detection sensitivity of the used techniques. This observation is in agreement with the low H-atom fluence used in the experiments and the presence of $H_2O$ ice which isolates the formed $CH_nO$ species from each other preventing their efficient interactions.

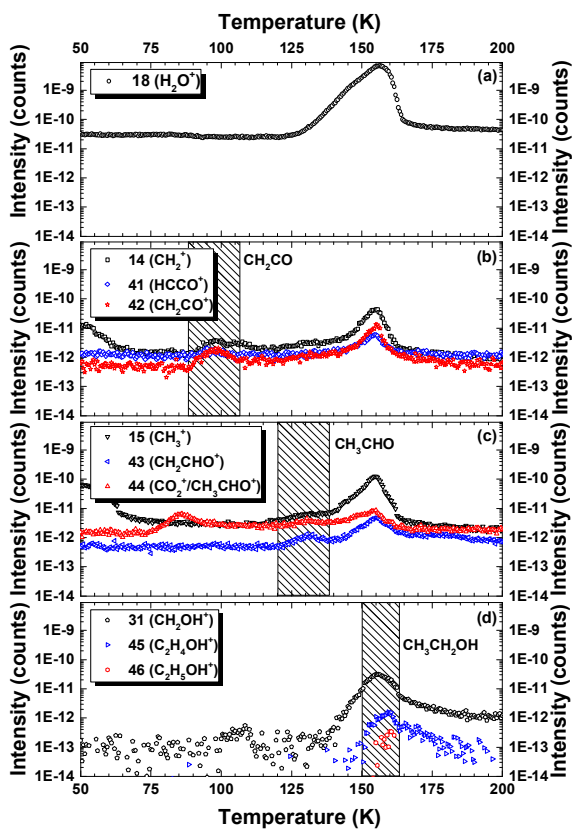

**Figure 2.** Four different mass settings for a TPD QMS experiment performed after the simultaneous co-deposition of $H_2O$ molecules with C atoms, CO molecules, and H atoms on a 10 K substrate for 1920 seconds. The $H_2O$:C:CO:H ratio is estimated to be 16:1:1.5:4 with a C-atom flux equal to $5 \times 10^{11}$ cm$^{-2}$. For clarity, the m/z plots are presented in distinct groups with characteristic m/z signals of (a) $H_2O$, (b) $CH_2CO$ (ketene), (c) $CH_3CHO$ (acetaldehyde), and (d) $CH_3CH_2OH$ (ethanol), respectively. The patterned boxes indicate the gas-phase detections of the various species at their characteristic desorption temperatures. The m/z=18 ($H_2O^+$) plot is shown to demonstrate that a vast fraction of the species formed, is expected to be entrapped in $H_2O$ ice and released into the gas-phase upon the sublimation of $H_2O$ ice. Spectra are presented without an offset for direct comparison.

### 3.3. C+$H_2CO$ and C+$CH_3OH$ reactivity

Along with the sequential hydrogenation of $CH_2CO$ yielding $CH_3CHO$ and $CH_3CH_2OH$ (Bisschop et al. 2007) as suggested here, a parallel reaction path may be involved. Insertion of C-atoms into the C-H bond of $H_2CO$ or $CH_3OH$ followed by hydrogenation of the formed species can also result in the formation of $CH_3CHO$ and $CH_3CH_2OH$. In line with the C-addition channel investigated here for ketene, such a process could also be relevant in the period just before the CO-freeze-out stage. For this reason, a series of additional experiments is performed to confirm or exclude this reaction mechanism. The experiment presented in spectrum (c) of Fig. 1 is repeated under the very same experimental conditions by using $H_2^{13}CO$ and $CH_3OH$ molecules instead of the admixture of $C^{18}O$. None of these experiments resulted in a clear identification of $CH_3CHO$ or $CH_3CH_2OH$ isotopes within the RAIRS sensitivity limits. This hints for the presence of activation barriers in $C+H_2CO$ and $C+CH_3OH$ reactions.

## 4. ASTROCHEMICAL IMPLICATIONS AND CONCLUSIONS

The experiments performed here aim at simulating chemical processes during the earliest stages of the star formation process. Upon transition from diffuse to translucent clouds, a fraction of elemental carbon previously locked in $C^+$ can be partially converted into neutral C-atoms before ultimately getting locked in molecular CO in dense molecular clouds. At this stage, the abundance of C-atoms with respect to elemental hydrogen can be as high as $10^{-4}$, see van Dishoeck & Black (1988), Snow & McCall (2006). This means that there is a period in the evolution of interstellar clouds when CO-molecules and C-atoms can both simultaneously accrete on grain surfaces. As this stage occurs early in or directly prior to the formation of the cold dark cloud core, it should coincide with the formation of the bulk of $H_2O$-ice produced by hydrogenation of accreting O-atoms (van Dishoeck et al. 2021). A few other main constituents of interstellar ices formed during this phase are simple hydrides $NH_3$ and $CH_4$, produced at least partially by hydrogenation of accreting N and C atoms, as well as $CO_2$ (Boogert et al. 2015, Linnartz et al. 2015). The latter one is produced from accreting CO molecules through interaction with OH radicals or direct addition of O atoms, see Chang & Herbst et al. (2012), Ioppolo et al. (2013). This stage directly precedes the so-called CO freeze-out stage, which implies higher densities and CO accretion rates.

In this study, the first experimental evidence for the formation of ketene on the surface of $H_2O$ ice at 10 K through interaction of CO molecules with C and H atoms is presented. Moreover, it is shown that the possible products of $CH_2CO$ hydrogenation, $CH_3CHO$, and traces of $CH_3CH_2OH$, form as well. Formation of $CH_2CO$ is expected to be initiated by $C+CO->:CCO$ interaction, which is reported to be barrierless, see Papakondylis & Mavridis (2019). The formed :CCO biradical, then, can be quickly hydrogenated into $CH_2CO$ by H-atoms by two consecutive barrierless radical-radical interactions, i.e., according to the reactions of the third scenario (Eqs. 3a-c) presented in the introduction.

To our knowledge, the hydrogenation of $CH_2CO$ forming $CH_3CHO$ at the specific conditions used in the present study was not demonstrated experimentally in the literature due to the difficulties associated with deposition of pure $CH_2CO$ ice. Gas-phase experiments reported a rather low barrier for this reaction, e.g., 975 K (Umemoto et al. 1984). This barrier is considerably lower than the effective activation barrier reported for CO hydrogenation for similar temperatures, meaning that it may efficiently proceed at 10 K (Andersson et al. 2011). Hydrogenation of pure $CH_3CHO$ into $CH_3CH_2OH$ under similar experimental conditions was studied in detail by Bisschop et al. (2007) and Chuang et al. (2020), and was shown to proceed at 10 K. However, an activation barrier for H-atom addition to a $CH_3CHO$ molecule was reported to be higher than that for CO hydrogenation, see Sivaramakrishnan et al. (2009). This means that formation of $CH_3CH_2OH$ from $CH_3CHO$ can be a rate limiting step, and therefore solid-state abundances of $CH_2CO$ and $CH_3CHO$ higher than that of $CH_3CH_2OH$ are expected early in the dark clouds.

Our suggested mechanism does not explicitly exclude other reaction routes, such as association of $:CH_2$ with CO (Eqs. 2a-c), see Krasnokutski et al. (2017). However, the presence of the 1500 - 1700 K activation barrier for $:CH_2$ + CO addition reaction in comparison to the barrierless C + CO interaction makes it an overall less probable mechanism (King et al. 1998; Ogihara et al. (2010), Papakondylis & Mavridis 2019). Similarly, C+CO reactivity appears to be a more probable mechanism than C + HĊO reactivity (Eq. 1b), as the latter involves interaction between two highly reactive radicals, while in the former case only C atoms can readily be consumed by surrounding species in the ice. Indeed, as HĊO radicals react with H atoms without the presence of an activation barrier, they have a significantly lower surface density than CO molecules, which are still required to bypass an activation barrier to be consumed in reaction with H atoms (see for example Fig. 3 of Aikawa et al. 2020, and Fig. 5 of Simons et al. 2020).

Following the successful identification of $CH_2CO$ here and evidence for the formation of $CH_3CHO$ and possibly $CH_3CH_2OH$ starting from ketene molecules, the previously suggested theoretical networks utilising solid-state C-atom addition reactions as the source of COM formation (see Charnley 2001) can get their first experimental confirmation. The updated formation routes, with inclusion of the results obtained in the work presented here, are shown in Figure 3. The reactions concluded from our work are marked with red and comprise the core of the C-atom addition/insertion network. The key step responsible for the growth of carbon skeleton and formation of 2-carbon bearing COMs is suggested to be the addition of a carbon atom to a CO molecule. The solid boxes connected with solid arrows sum-up other related and well-known reaction routes investigated previously both theoretically and in the laboratories, see Linnartz et al. (2015) and references therein. The grey dotted boxes and arrows suggest some of the speculative routes, which may result in **a** further increase of the chemical complexity and formation of 3-carbon bearing derivatives. These routes imply C-atom addition to the unsaturated carbon-carbon bonds followed by consequent hydrogenation of the formed reactive intermediates (Kaiser & Mebel 2002, Henning & Krasnokutski 2019) and are in line with the ideas initially presented by Charnley and co-workers. Dedicated future work, combining experimental and theoretical approaches, will be required to fully pin-point the relative impact for such C-atom induced solid state reaction routes, i.e., when comparing with pathways involving recombination of $CH_nO$ radicals or addition of OH radicals to the unsaturated CC bonds of hydrocarbons (Chuang et al. 2016, 2020, Qasim et al. 2019).

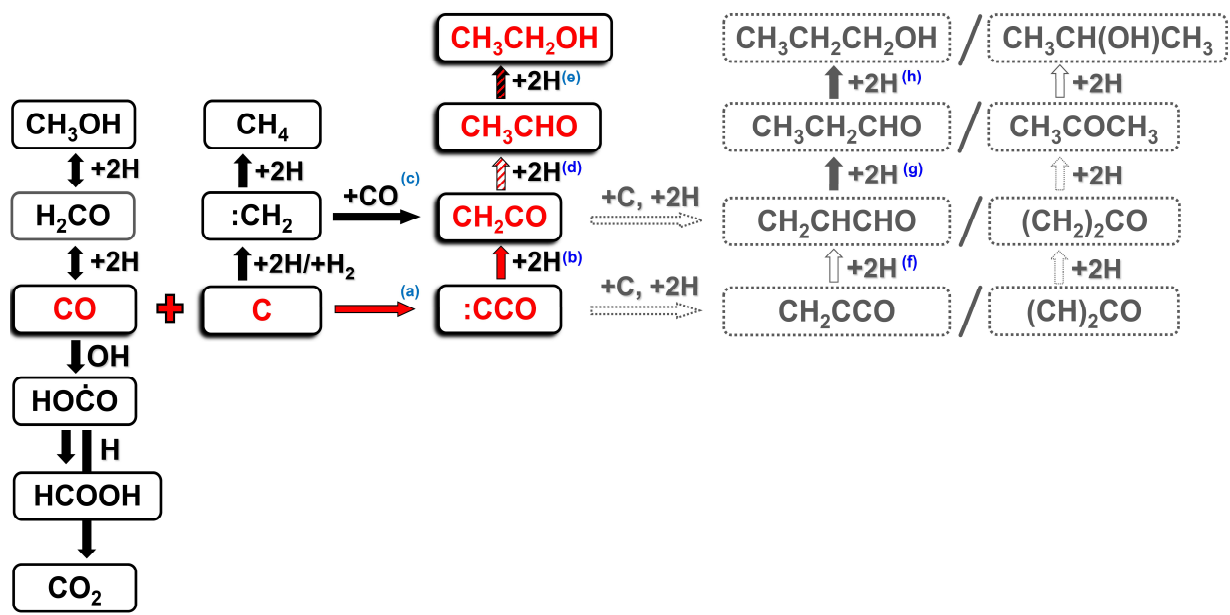

**Figure 3.** Proposed COM formation network initiated by C-atom addition to CO molecules. The reaction routes investigated in the present study are marked in red. Filled arrows indicate experimentally verified formation routes. Empty arrows indicate reaction routes predicted theoretically. Grey dotted arrows and boxes account for the speculative routes resulting in the formation of 3-carbon bearing derivatives. These routes are suggested in view of the efficient reactivity of C atoms with the unsaturated carbon-carbon bonds (Kaiser & Mebel 2002, Henning & Krasnokutski 2019). $(CH)_2CO$ and $(CH_2)_2CO$ stand for cyclopropenone and cyclopropanone, respectively.

---

The activation barriers for the rate-determining reactions obtained in this work or from the literature:

[a] 0 K, Papakondylis & Mavridis 2019;

[b] 0 K, H+CCO and H+HCCO are expected to proceed barrierlessly as radical-radical interaction;

[c] 1500 – 1700 K, King et al. 1998, Ogihara et al. 2010;

[d] 975 K, Umemoto et al. 1984;

[e] 3570 K, Sivaramakrishnan et al 2009;

[f] 0 K, Shingledecker et al 2019;

[g] 990 K, Zaverkin et al 2018;

[h] 2900 K, Zaverkin et al 2018.

Efficient solid-state formation of $CH_2CO$, $CH_3CHO$, and $CH_3CH_2OH$ initiated by accretion of C-atoms, H-atoms, and CO-molecules means that these COMs are likely already formed on icy grains early in the lifetime of interstellar clouds, i.e., well before the occurrence of CO freeze-out stage commonly associated with formation of a large fraction of COMs described with formula $(CO)_nH_m$. The occurrence of C-atom induced solid-state COM formation routes makes $CH_3CHO$ and $CH_3CH_2OH$ promising candidates for upcoming JWST observations, as they can be formed both during the early dark cloud evolution and later upon the processing of CO-rich ice containing $CH_3OH$ by various chemical triggers, thus, higher total column densities of $CH_3CHO$ and $CH_3CH_2OH$ should be available for the observations (Öberg et al. 2009, Modica & Palumbo 2010, Vasyunin & Herbst 2013, Maity et al. 2015, Fedoseev et al. 2017). The abundance of $CH_3CHO$ and $CH_3CH_2OH$ prior to the CO freeze-out can be further enhanced by the $C_2/C_2H_n$-induced formation route presented in Bergner et al. (2019) and Chuang et al. (2020), (2021). Indeed, $CH_3CHO$ and $CH_3CH_2OH$ are a few of the COMs tentatively assigned in the solid state as the possible carriers of the 7.41 and 7.24 µm ice bands, respectively (Schutte et al. 1999; Öberg et al. 2011, Terwisscha van Scheltinga et al. 2018). The distinct profiles of $CH_3CHO$ and $CH_3CH_2OH$ absorption features in $H_2O$-rich and $CO:CH_3OH$-rich ices, see Terwisscha van Scheltinga et al. (2018), may also provide insights on the typical environment in which $CH_3CHO$ and $CH_3CH_2OH$ molecules are embedded, linking the formation of these molecules to specific star-formation stages.

The formed $CH_2CO$ and $CH_3CHO$, in turn, can be partially transferred from the solid-state into the gas-phase by various thermal and non-thermal mechanisms contributing to the gas-phase abundances of these species in dark and translucent clouds (Dulieu et al. 2013, Rawlings et al. 2013, Chuang et al. 2018, Dartois et al. 2019, Hoang & Tram 2020). Enrichment of the gas-phase by $CH_2CO$ and $CH_3CHO$ or by intermediate products obtained along their solid-state formation routes may provide the initial reagents for otherwise inefficient gas-phase reactions. This can further enhance the gas-phase composition during the earliest stages of star-formation (Vasyunin & Herbst 2013). For the nearby future, detailed astrochemical simulations utilising gas-grain models are required to estimate the exact impact of the verified formation routes on the gas-phase and solid-state abundances of $CH_2CO$, $CH_3CHO$, and $CH_3CH_2OH$, providing more information on the relative amounts of these species during the earlier and later stages of star formation.

**Acknowledgement**

This research work is funded by DANII (Dutch Astrochemistry Network II) and NOVA (the Netherlands Research School for Astronomy). The work extends on financial support obtained through a VICI grant from NWO (the Netherlands Organization for Scientific Research) and A-ERC grant 291141 CHEMPLAN. We are also grateful for financial support from the Danish National Research Foundation through the Center of Excellence 'InterCat' (Grant agreement no. DNRF150). G.F. acknowledges financial support from the Russian Ministry of Science and Higher Education via the State Assignment Contract FEUZ-2020-0038. S.I. acknowledges the Royal Society for financial support. T.L. is grateful for support from NWO via a VENI fellowship (722.017.008). The described work has benefited a lot from collaborations within the framework of the FP7 ITN LASSIE consortium (GA238258). We are very grateful to T. Banu and J. Kästner for sharing the results of their independent calculations on the activation barrier of C+CO reaction. We thank J. He for stimulating discussions and would like to express special gratitude to S. Krasnokutski for sharing his valuable experience in operating of the C-atom source.

**APPENDIX A.**

The RAIR spectra presented in Figure 1 from 2250 to 1200 cm$^{-1}$ are shown in Figure A.1 for the full recorded range from 4000 to 700 cm$^{-1}$. Several additional assignments can be made: CH$_4$ ($\nu_1$, 3009 cm$^{-1}$) and CD$_4$ ($\nu_3$, 994 cm$^{-1}$), see Chapados & Cabana 1972, and the broad H$_2$O absorption bands peaking around 3390 cm$^{-1}$ and 820 cm$^{-1}$, see Gerakines et al. 1995. The absorption feature of solid state CO$_2$ ($\nu_3$, 2343 cm$^{-1}$) is in part concealed by the absorption feature of gas-phase CO$_2$ present in the purging gas along the path of the IR beam outside of the setup. The weaker absorption feature in the range from 3700 to 3600 cm-1, in the shoulder of the strong 3390 cm$^{-1}$ H$_2$O band, is assigned to the dangling OH vibrations of H$_2$O molecules, see Sadlej et al. 1994. The increase in the absorption of dangling OH mode of H$_2$O molecules in the spectrum (c) is consistent with the addition of C$^{18}$O molecules to the co-deposition experiment and increased separation between H$_2$O molecules in the bulk of the mixed ice.

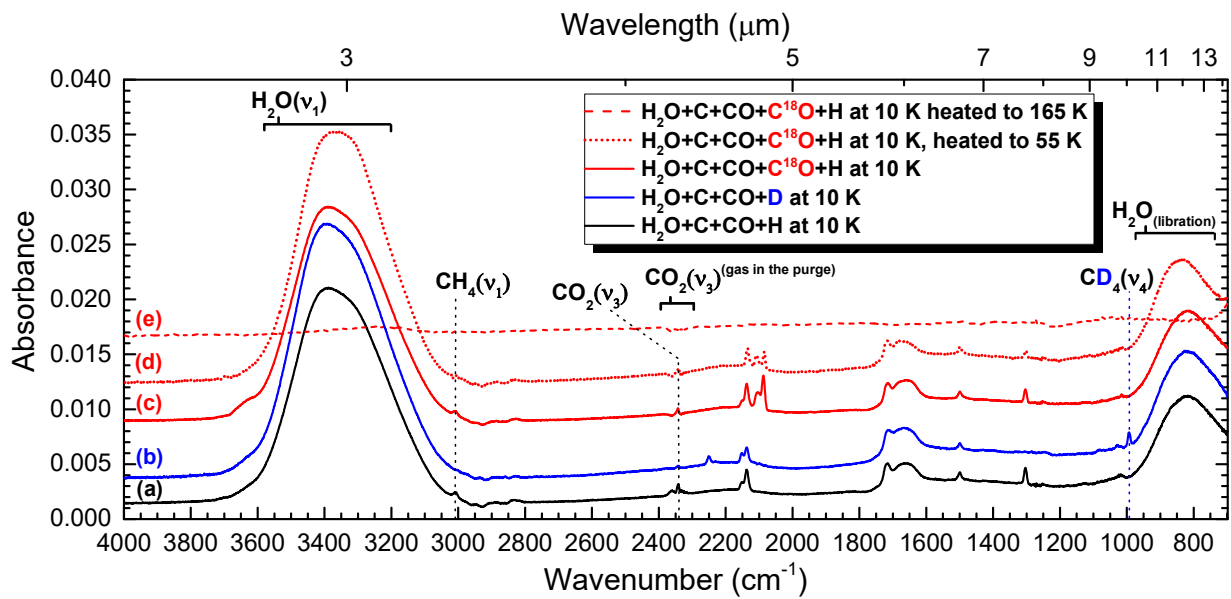

**Figure A.1.** The full range IR spectra (4000–700 cm$^{-1}$) of the RAIRS signals shown in Fig. 1. Spectra are offset for clarity.

**References**

Abplanalp, M. J., Förstel, M., Kaiser, R. I. 2016, Chem. Phys. Lett., 644, 79
Aikawa, Y., Furuya, K., Yamamoto, S., Sakai, N. 2020, Astrophys. J., 897, 110
Albar, J. D., Summerfield, A., Cheng, T. S., et al. 2017, Sci. Rep., 7, 6598